\documentclass{jetpl}

\usepackage{cite}
\twocolumn

\lat

\title{Density bump formation in a collisionless electrostatic shock wave in a laser-ablated plasma}

\rtitle{Density bump formation in a collisionless electrostatic shock wave in a laser-ablated plasma}

\sodtitle{Density bump formation in a collisionless electrostatic shock wave in a laser-ablated plasma}

\author{M.\,A.\,Garasev$^{a,b}$, A.\,I.\,Korytin$^{a}$, V.\,V.\,Kocharovsky$^{a}$, Yu.\,A.\,Malkov$^{a}$, A.\,A.\,Murzanev$^{a}$, A.\,A.\,Nechaev\/\thanks{ant.a.nech@gmail.com}$^{a}$, A.\,N.\,Stepanov$^{a}$}

\rauthor{M.\,A.\,Garasev, A.\,I.\,Korytin, V.\,V.\,Kocharovsky, Yu.\,A.\,Malkov, A.\,A.\,Murzanev, A.\,A.\,Nechaev, A.\,N.\,Stepanov}

\sodauthor{Garasev, Korytin, Kocharovsky, Malkov, Murzanev, Nechaev, Stepanov}

\address{$^a$ Institute of Applied Physics RAS,
603950 Nizhny Novgorod, Russia\\~\\
$^b$ Lobachevsky State University of Nizhny Novgorod, 603950 Nizhny Novgorod, Russia}

\dates{\today}{*}

\abstract{
The emergence of a density bump at the front of a collisionless electrostatic shock wave have been observed experimentally during the ablation of an aluminium foil by a femtosecond laser pulse. 
We have performed numerical simulations of the dynamics of this phenomena developing alongside the generation of a package of ion-acoustic waves, exposed to a continual flow of energetic electrons, in a collisionless plasma.
We present the physical interpretation of the observed effects and show that the bump consists of transit particles, namely, the accelerated ions from the dense plasma layer, and the ions from the diluted background plasma, formed by a nanosecond laser prepulse during the ablation.
}


\begin{document}

\maketitle
{\bf 1. Introduction.}
Quasielectrostatic shocks in a nonequilibrium plasma, formed due to the charge separation of hot and cold fractions, have been studied theoretically and experimentally for more than half a century \cite{bib_1, bib_2, bib_3, bib_4, bib_SARRI, bib_c1, bib_MED}.
One of the most studied classes of such waves are ion-acoustic shocks, which can occur during an explosive process in plasma and after the formation stage are described by the theory of ion-acoustic solitons \cite{bib_5, bib_6, bib_x}. 
The example of such an explosive process is the ablation of a solid target by nano- or picosecond laser pulses, which leads to the formation of an electrical double layer and the excitation of a short package of ion-acoustic waves \cite{bib_c3, bib_STEP, bib_7, bib_8, bib_9}. 
In many experiments this package, while propagating in surrounding plasma, is affected by a fast flow of nonequilibrium particles or the laser field for a long time after the explosion. 
The dynamics of the shock wave at this stage acquires a number of peculiarities that are not described by the soliton theory and are associated with the complex collisionless behavior of particles. 
This dynamic remains experimentally unexplored (see \cite{bib_c1, bib_8, bib_9, bib_9x}).

In this letter we explore the formation and evolution of a density bump in such shock waves and study the collisionless kinetics of ions and electrons in the bump during its propagation.
We reveal the crucial role of the background plasma, formed by laser radiation prior to main pulse, in the process of a bump formation (the existing approximate solutions to the problem of the plasma expansion into vacuum demonstrate no bumps; see, for example, \cite{bib_10, bib_11, bib_12}).

\begin{figure}[t]
	\includegraphics[width=0.50\textwidth]{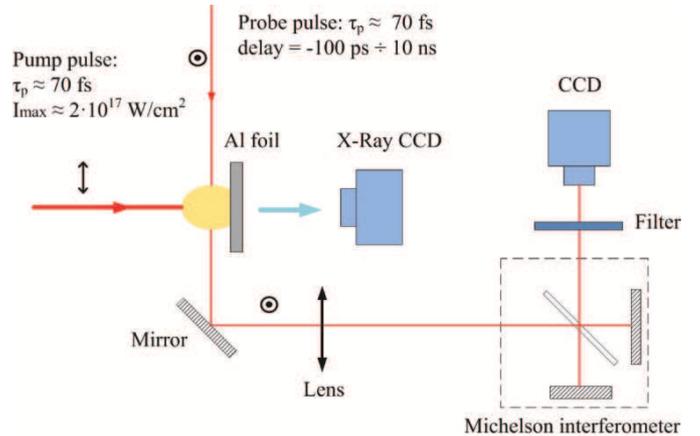}
	\caption{Fig. 1. Experimental setup}
	\label{fig1}
\end{figure}

{\bf 2. Experimental setup.}
In experiments (see Fig.~\ref{fig1}) we used a Ti:sapphire femtosecond laser system with a pulse energy of $160$~mJ, a pulse duration $\tau_\mathrm {p} = 70$\,fs, emission wavelength of $800$\,nm, and a repetition rate of $10$\,Hz. 
The laser beam was focused on the flat surface of a $200$\,$\mu$m-thick aluminium foil with the help of a $50$\,cm spherical mirror. 
The beam size at the focus was $d \approx 40-50$~$\mu$m and the maximum intensity reached $2\times10^{17}$~W\,cm$^{-2}$.
Experiments were carried out in the vacuum chamber with a residual gas pressure less than $0.05$~Torr.
The power contrast of pulses was measured with a single-pulse autocorrelator for times up to $\sim1$~ps and with a $5$\,GHz photodiode for times larger than 200~ps. 
The contrast for times up to $1$~ps was $\sim10^{-3}$, the value of the contrast with respect to the amplified spontaneous emission (arriving $2-3$~ns before the main pulse) was estimated as $(2-5)\times 10^{-7}$.

Diagnostics of the plasma density was carried out with the use of a Michelson interferometer.
The scan pulse, separated from the main laser beam, crossed the plasma perpendicular to the axis of the pump pulse's propagation at an adjustable time delay. 
The image from the interferometer's output was recorded by a CCD camera. 
Interferograms had the spatial resolution of $5$~$\mu$m with a field of view of $\sim 1$~mm$^2$. 
To reduce the background light, the polarization of the diagnostic beam was chosen to be orthogonal to the polarization of the pump.
In front of the CCD camera a bandpass interference filter ($785\pm20$~nm, Edmund Optics) was installed to increase the coherence length of the interferometric images. 
The X-ray CCD camera (Andor DO 3040), placed behind the target, allowed us to measure the spectrum of bremsstrahlung radiation of the plasma and to determine the electron temperature of the latter.

\begin{figure}[h]
	\includegraphics[width=0.50\textwidth]{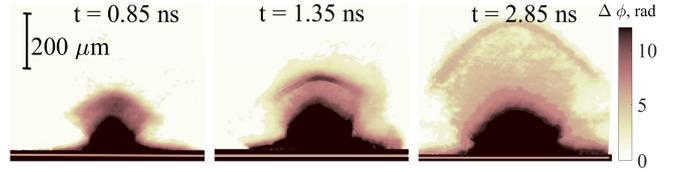}
	\caption{Fig.~2. The spatial distribution of the phase shift, caused by the plasma, for different time delays with respect to the pump pulse.}
	\label{fig2}
\end{figure}

{\bf 3. Observed parameters of the plasma and the shock wave.}
The spatial distribution of the phase shift, caused by the plasma produced by a femtosecond pulse, is shown in Fig.~\ref{fig2} for different time delays of the probe.
About 100~ps before the main pulse arrives to the target, a preplasma formed by a nanosecond prepulse exists near the foil's surface, having a size of $\sim$50~$\mu$m and a number density of the order of $10^{19}$\,cm$^{-3}$.
When the pump hits the target (at $t=0$), a multiple increase of the total amount of ejected ionized material occurs.
In the experiment the electron temperature of the plasma, estimated from the measurements of X-ray bremmstrauhlung spectrum, was about $2-2.5$~keV.

\begin{figure}[h]
	\includegraphics[width=0.5\textwidth]{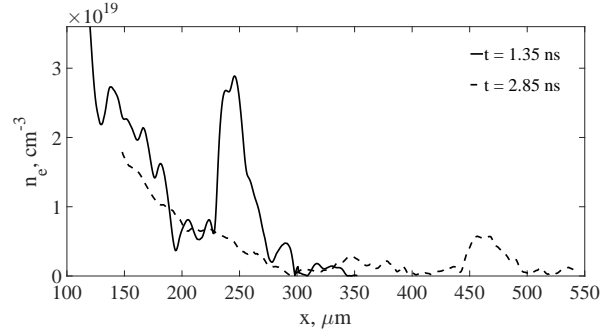}
	\caption{Fig.~3. Two snapshots of the plasma density along the axis of expansion.}
	\label{fig3}
\end{figure}

As seen from Fig.~\ref{fig2}, the essential feature of the spatial distribution of the number density is the existense of a bump at the front edge of the expanding plasma.
Fig.~\ref{fig3} shows the number density distribution along the middle line of the plasma flow for $t = 1.35$~ns and $t = 2.85$~ns, obtained applying the inverse Abelian transformation to the interferometric data.
The clearly visible bump holds the electron number density of $2\times10^{19}$cm$^{-3}$ and is followed by the plasma tail expanding from the surface of the foil.
The bump propagates at the speed of about $v\sim1.5\times10^7$cm s$^{-1}$, virtually constant on the times up to $t_{max} = 3.5$~ns.
Note that, although laser ablation processes, including the formation of the surface plasma and the expanding plasma, have already been considered in several works (e.g., \cite{bib_c1, bib_c3, bib_STEP, bib_7, bib_9x}), the observed phenomenon of a bump formation at the front of a collisionless electrostatic shock wave is practically unexplored.

{\bf 4. PIC-simulation: Formulation of the problem and initial setup.}
Estimates show that the experimentally observed shock exists in almost collisionless plasma. 
Indeed, at a number density $n_e\sim10^{20}$\,cm$^{- 3}$ and a temperature $T\sim 2.5$~keV, typical for the experiments, the mean free path of electrons is about $300$\,$\mu$m, thus greater than the size or the region where the observed shock wave formation happens.
At times under consideration, $t\lesssim3$~ns, the heated area on the foil's surface and the compact quasi-adiabatically expanding plasmoid, with a number density of $\gtrsim 10^{21}$cm$^{-3}$, act as a quasi-stationary source of the mentioned collisionless plasma.
It constantly emits hot electrons (a part of which returns) with energies of $\sim 2.5$~keV that pull ions and transmit energy to them through a quasi-electrostatic field (see, e.g., \cite{bib_STEP}).

Having no aim to give a complete description of the dynamics of a dense plasma created by a powerful laser pulse, we have performed 1D and 2D numerical simulations of the expansion of its outer layers, using relativistic fully kinetic PIC (particle-in-cell) code EPOCH.
The code consistently solves Vlasov's equations for the plasma particles' motion with Maxwell's equations for electromagnetic fields \cite{bib_EPOCH}.

\begin{figure}[h]
	\includegraphics[width=0.50\textwidth]{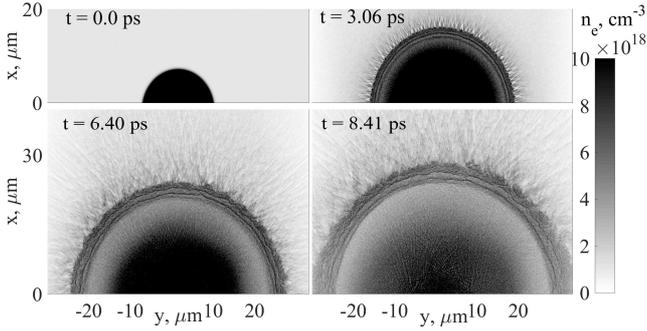}
	\caption{Fig. 4. Snapshots from a 2D simulation showing the formation of a bump in plasma number density. 
Simulation parameters: $n_\mathrm{bkg} = 10^{18}$\,cm$^{-3}$, $ n_\mathrm{L} = 10^{20}$\,cm$^{-3}$, $ M/m_e =  100$, ion charge $Z = 1$. }
	\label{fig4}
\end{figure}

In 1D geometry as the initial configuration for a simulation we took cold (neutral) background plasma with the electron density $n_\mathrm{bkg} = 10^{15}-10^{19}$cm$^{- 3}$ (for different runs) that filled the entire computational domain, and a compact (also neutral) plasma layer, placed near the left boundary of the domain and occupying one sixth of its length, which constituted cold ions with charge $Z$ and hot electrons with the number density $ n_\mathrm{L} = 10^{20}$\,cm$^{-3}$.
The transition profile from the layer to the background was super-Gaussian, $n_\mathrm{L}\sim\mathrm{e}^{-x^\ell/L^\ell}$, where $L$ is the effective size of the layer and $x$ is the coordinate along the direction of expansion.
In this report all the results presented in figures are given for the case of the sharp profile with $\ell = 8$.
On the left boundary of the domain we used a reflective condition, the right boundary was open (absorbing). 
The ion-electron mass ratio varied within the range $M/m_e = 100-50 000$ (for different runs). 
Initially all the particles in the simulation had an isotropic Maxwellian velocity distribution; the electron temperature in the layer was $T_\mathrm{L} = 2.5 $~keV, background electrons and all the ions had $T_\mathrm{bkg} = 3$~eV.

In two-dimensional calculations we used similar initial conditions, except that the dense plasma had the shape of a half a circle, still denoted here as "the layer" though; lateral boundaries were open.

\begin{figure}[h]
	\includegraphics[width=0.50\textwidth]{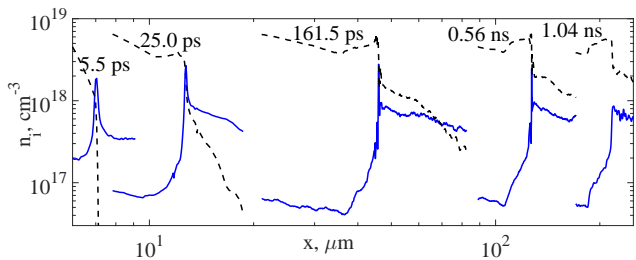}
	\caption{Fig. 5. Profiles of ion number density at consequent time-moments, obtained in a 1D simulation of the plasma expansion with parameters $Zn_\mathrm{i, bkg} = 10^{18}$\,cm$^{- 3}$, $ Zn_\mathrm{i, L} = 10^{20}$\,cm$^{- 3}$, $M/m_e = 50000$, ion charge $Z = 3$. 
Solid (blue) curves are for the background ion number density $n_\mathrm{i, bkg}$, dashed (black) curves — the number density of ions from the expanding layer $n_\mathrm{i, L}$.
The speed of the bump is $v_\mathrm{bump}\approx (1.8-2.1) \times 10^7$\,cm\,s$^{-1}$, being in a good agreement with experimentally measured values.}
	\label{fig5}
\end{figure}

The temporal evolution of the plasma number density is shown in Fig.~\ref{fig4}, Fig.~\ref{fig5} for 2D and 1D simulations respectively. 
Qualitatively 1D and 2D results are consistent with each other and reproduce well the situation observed in the experiment at the transitional stage of the shock wave's formation, when a continuous flow of energetic electrons from the hot plasma layer is present.

\begin{figure}[t]
	\includegraphics[width=0.50\textwidth]{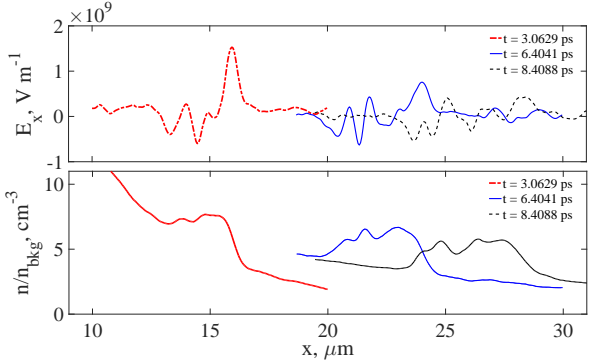}
	\caption{Fig. 6. Profiles of the electric field and the number density near the bump along the $x$-axis for the 2D run shown in Fig. 4.}
	\label{fig6}
\end{figure}

{\bf 5. Dynamics and kinetics of the density bump's formation.} 
At the initial stage the plasma expansion is actually one-dimensional: the layer of hot electrons rushes forward, leaving ions behind.
At times less than or about one plasma period $\tau \equiv 2 \pi / \omega_\mathrm{pl} =\sqrt{\pi m_e/ (e^2n_\mathrm{L})}$ the region that holds the point of the maximum of the growing longitudinal electric field $E_x (t, x)$ moves at the speed of the order of the thermal velocity of electrons.
After $(1-2)\tau$ the nonlinear stage begins: the electric field significantly slows down electrons and begins to accelerate ions.
At about $3\tau$ (for the parameters indicated in Fig.~\ref{fig4}) the value of the electric field reaches its maximum, and then begins to decrease.
After $\sim 50\tau$ the speed of the field maximum desreases to the approximately ion-acoustic one $v_\mathrm {is} = \sqrt{ZT_\mathrm{L} / M}$.
%
%
At this point in the spatial profile of $E_x (x)$ a quasi-stationary structure is formed (see Fig.~\ref{fig6}, top panel). It has a width of $\sim10$ Debye lengths, holds several field oscillations and moves at almost constant velocity, slightly above $v_\mathrm{is}$ (Mach number $ \approx 1.2-1.4 $).
During the propagation this wave package broadens only slightly, no more than twice on the time scales considered in the simulations (up to several thousands of $\tau$ in 2D runs, hundreds of thousands in 1D).
%
%
This structure's field accelerates the overtaken background ions up to velocities of the order of its own. 
The number density profile of both electrons and ions in this structure has a local maximum — the density bump (Fig.~\ref{fig6}, bottom panel). 
At the same time, in the plasma layer's part that remains behind the bump and is depleted of electrons, a smoothly oscillating electric field with a large total drop of potential arises, accelerating ions of the layer up to velocities several times greater than the ion-acoustic one.

The bump's formation occurred qualitatively the same way for different values of the ion to electron mass ratio, which varied in simulations from 100 to 50000. However, a considerable excess of the maximum number density at the shock front over the minimum behind it is observed only for a sufficient number density of the background plasma. 
The density bump was barely noticeable and quickly dissapeared (within a few tens of $\tau$), if the layer to background plasma density ratio was significantly less than $10^{- 2}$.
In general, we can state that the backround ions are responsible most of all for the emergence of the local maximum in plasma number density, and that the ions should be quite cold. 
Our simulations showed that the whole process is almost independent either of the layer ion temperature, as long as it is less than or of the order of its electron temperature, or the temperature of the background electrons (because of their insignificant density).

\begin{figure}[t]
	\includegraphics[width=0.50\textwidth]{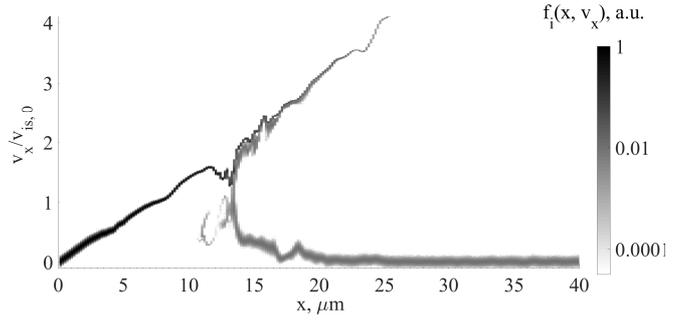}
	\caption{Fig. 7. The distribution function of an ion velocity and longitudinal coordinate.
The upper inclined curve, including the part before the branching, mainly consists of the layer ions, while the lower one corresponds to the background.}
	\label{fig7}
\end{figure}

It is very informative to investigate the evolution of the distribution of ions from the layer and the background in the longitudinal velocity - coordinate space (Fig.~\ref{fig7}).
According to \cite{bib_SARRI}, the appearance of branching features in this phase plane is an indicator of a shock wave presence in the system. 
The jump of the mean velocity associated with it, i.e. the ratio of ion mean speeds before and behind the shock front, quickly reaches the value of about $1.2$ and then remains approximately constant.
The velocity of the density bump is also almost constant, despite the fact that over time the average energy of electrons decreases more than twofold.
During the shock wave's formation, the background ions are hardly reflected by the shock, rather being accelerated by its field or partially trapped, forming a density bump at its front.
%
%
Ions from the dense layer can be accelerated to greater speeds through the electric field behind the shock, so they do not only increase the plasma density in the bump, but also overtake the shock wave, gradually changing and smoothing its front.
The relative contribution of ions from the layer in the bump density constantly increases with time, compared to the contribution of the background ions, and soon after the bump's formation begins to dominate (see Fig.~\ref{fig5}). 
It is interesting to note, that the background ions are concentrated in the narrow layer in the front part of the bump. 
The ratio of the maximum density in the bump to that before the bump and to the minimum density behind it (considering the width of about $\sim10$ Debye lengths) rapidly increases to values of $\sim 3.5$ and $\sim 1.6$ respectively and then slowly decreases. Note, that during the simulation time, the average thermal energy of electrons is reduced more than twice.

It should be noted, that both ion fractions are not trapped in the bump for long, and either overtake it or lag behind, making sometimes one or two oscillations. 
As a result, behind the front streams of ions are formed that move at different velocities and thus increase the width of the shock front and the number of field oscillations in it.
Despite this, the shape of the density bump, which is situated at the front of the shock, remains almost constant for a long time, as well as its velocity, the total energy and the number of particles in it. 
The ratio of the energy of the particles in the bump to the initial energy of the hot plasma layer particles for the chosen initial conditions (see. Fig.~\ref{fig4}) reaches 5\%.

The existence of such a density bump and an electrostatic shock wave as a whole ows to the pressure associated with the gradient of the energetic electrons number density and is consistent with the drop of that pressure at the shock front. 	
Сomprehensive analysis of the long-term evolution of such a "forced" shock wave and a bump in it, as well as the possible formation of freely propagating ion-acoustic solitons, requires taking into account the significant depletion of said energetic electrons pressure and is beyond the scope of this report.
In typical conditions this evolution is accompanied by the development of a Weibel instability and the generation of quasi-static magnetic fields that could affect both the overall redistribution of energy between electrons and ions, and the mixing of particles' trajectories behind the shock front. 
A detailed description of mechanisms of the magnetic field generation, which energy could reach $\sim10$\% of the initial energy of the layer particles, will be the subject of a separate work.

The dependence of the density bump's shape and the structure of the supporting electric field on the initial shape of the transition region between the dense and the background plasma also deserves a special study.
We discovered such a dependency for power-law density profiles, although it is absent for sharp super-Gaussian profiles discussed here.

{\bf 6. Conclusion.}
In summary, both the experiments on femtosecond laser ablation and the numerical PIC-simulations indicate that the occurrence of a plasma density bump at the front of an electrostatic collisionless shock are the result of an explosive heating of plasma electrons and their subsequent pressure on the generated package of ion-acoustic waves. 
The necessary condition for the development of the bump is the presence of a not too diluted plasma background.
Further investigation of the processes of formation and evolution of these forced shock waves in laser plasma experiments are of great interest.

This work was supported in part by the Government of the Russian Federation Project No.
14Z50.31.0007 (sections 2-4 of the letter) and the Russian Science
Foundation grant No 16-12-10528 (section 5).


\begin{thebibliography}{9}
\bibitem{bib_1}
S.\,S. Moiseev and R.\,Z. Sagdeev, J. Nucl. Energy {\bf 5}, 43 (1963).

\bibitem{bib_2}
V.\,L. Krasovsky, H. Matsumoto, and Y. Omura, Journal of Geophysical Research {\bf 108}, 1117 (2003).

\bibitem{bib_3}
A. Balogh, R. Treumann, Physics of Collisionless Shocks, Springer (2003).

\bibitem{bib_4}
C. M Huntington, F. Fiuza, J.S. Ross, A.\,B. Zylstra, R.\,P. Drake, D.\,H. Froula, G. Gregori, N.\,L. Kugland, C.\,C. Kuranz, M.\,C. Levy, C.\,K. Li, J. Meinecke, T. Morita, R. Petrasso, C. Plechaty, B.\,A. Remington, D.\,D. Ryutov, Y. Sakawa, A. Spitkovsky, H. Takabe, H.-S. Park, Nature Physics, {\bf 11}, 173 (2015)

\bibitem{bib_SARRI}
G. Sarri, G.\,C. Murphy, M.\,E. Dieckmann, A. Bret, K. Quinn, I. Kourakis, M. Borghesi, L.\,O.\,C. Drury, A. Ynnerman,  New Journal of Physics, {\bf 13}, 073023 (2011)

\bibitem{bib_c1}
H. Ahmed, M.\,E. Dieckmann, L. Romagnani, D. Doria, G. Sarri, M. Cerchez, E. Ianni, I. Kourakis, A.\,L. Giesecke, M. Notley, R. Prasad, K. Quinn, O. Willi, M. Borghesi,. Phys. Rev. Lett {\bf 110}, 205001 (2013)

\bibitem{bib_MED}
Yu.\,V. Medvedev, Nonlinear features of the decay of discontinuities in rarefied plasma (in Russian), Moscow: Fizmatlit (2012)

\bibitem{bib_5}
L.\,A. Artsimovich, R.\,Z. Sagdeev,  Plasma physics for physicists (in Russian),  Moscow: Atomizdat, (1979). 

\bibitem{bib_6}
S. Sultana, G. Sarri, I. Kourakis,  Physics of Plasmas, {\bf 19}, 012310 (2012)

\bibitem{bib_x}
E.\,J. Choi, K. Min, K.-I. Nishikawa, C.\,R. Choi, Physics of Plasmas, {\bf 21}, 072905 (2014)

\bibitem{bib_c3}
T. -H. Tan, J.\,E. Borovsky, J. Plasma Physics, {\bf 35}, 239 (1986)

\bibitem{bib_STEP}
O.\,B. Anan'in, Y.\,V. Afansiev, Y.\,A. Bykovsky, O.\,N. Krokhin, Laser plasma (in Russian), Moscow: MEPHI (2003)

\bibitem{bib_7}
 N.\,C. Woolsey, Y. Abou Ali,  R.\,G. Evans, R.\,A.\,D Grundy, S.\,J. Pestehe, P.\,G. Carolan, N.\,J. Conway, R.\,O. Dendy, P. Helander, K.\,G. McClements, J.\,G. Kirk, P.\,A. Norreys, M.\,M. Notley, S.\,J. Rose, Physics of Plasmas, {\bf 8},  2439 (2001)

\bibitem{bib_8}
L. Romagnani, S.\,V. Bulanov, M. Borghesi, P. Audebert, J.\,C. Gauthier, K. L\"{o}wenbr\"{u}ck, A.\,J. MacKinnon, P. Patel, G. Pretzler, T. Toncian, O. Willi, Phys. Rev. Lett, {\bf 101}, 025004 (2008) 

\bibitem{bib_9}
C. Ruyer, L. Gremillet, G. Bonnaud, Physics of Plasmas, {\bf 22}, 082107 (2015)

\bibitem{bib_9x}
K.\,A. Ivanov, S.\,A. Shulyapov, P.\,A. Ksenofontov, I.\,N. Tsymbalov, R.\,V. Volkov, A.\,B. Savel'ev, A.\,V. Brantov, V.\,Yu. Bychenkov, A.\,A. Turinge, A.\,M. Lapik, A.\,V. Rusakov, R.\,M. Djilkibaev, V.\,G. Nedorezov, Physics of Plasmas, {\bf 21}, 093110 (2014)

\bibitem{bib_10}
D.\,S. Dorozhkina, V.\,E. Semenov, JETPL, {\bf 67},  573 (1998)

\bibitem{bib_11}
D.\,S. Dorozhkina, V.\,E. Semenov, JETP, {\bf 89}, 468 (1999)

\bibitem{bib_12}
P. Mora, T. Grismayer, Phys. Rev. Lett, {\bf 102},  145001 (2009)

\bibitem{bib_EPOCH}
T.\,D. Arber, K. Bennett, C.\,S. Brady, A. Lawrence-Douglas, M.\,G. Ramsay, N.\,J. Sircombe, P. Gillies, R.\,G. Evans, H. Schmitz, A.\,R. Bell, C.\,P. Ridgers, Plasma Physics and Controlled Fusion, {\bf 57}, 113001 (2015)

\end{thebibliography}
\end{document}